\documentclass[twocolumn,prl,twocolumn,showpacs,amsmath,amssymb,preprintnumbers]{revtex4}
\usepackage{eprint}
\usepackage{hyperref}
\usepackage{graphicx}
\usepackage{dcolumn}% Align table columns on decimal point
\usepackage{bm}% bold math

\usepackage{amsthm}
\newtheorem*{conj}{Conjecture}

\newcommand{\prob}[1]{\mathbb{P}_Q\!\left( #1 \right)}
\newcommand{\probL}[1]{\mathbb{P}_L\!\left( #1 \right)}
\newcommand{\pr}[1]{\mathbb{P}\!\left( #1 \right)}

\newcommand{\ket}[1]{ | #1 \rangle}
\newcommand{\bra}[1]{ \langle #1 |}

\newcommand{\Ac}{\mathcal{A}}

\newcommand{\la}{\lambda}
\newcommand{\C}{\mathbb{C}}

\def\one{\leavevmode\hbox{\small1\normalsize\kern-.33em1}}

\begin{document}

\title{On the maximal violation of the Collins-Gisin-Linden-Massar-Popescu inequality for infinite dimensional states}

\author{Stefan Zohren$^{1}$}
\author{Richard D. Gill$^{2}$}
\affiliation{${}^1$Mathematical Institute, Utrecht University and Blackett Laboratory, Imperial College London\\
${}^2$Mathematical Institute, University of Leiden and EURANDOM, Eindhoven}

\date{\today}
\preprint{IMPERIAL-TP-06-SZ-05}

\begin{abstract}
We present a much simplified version of the CGLMP inequality for the $2\times2\times d$ Bell scenario. Numerical maximization of the violation of this inequality over all states and measurements suggests that the optimal state is far from maximally entangled, while the best measurements are the same as 
conjectured best measurements for the maximally entangled state. 
For very large values of $d$ the inequality seems to reach its minimal value given by the probability constraints.
This gives numerical evidence for a tight quantum Bell inequality
(or generalized Csirelson inequality) for the $2\times2\times\infty$ scenario.
\end{abstract}
\maketitle

%\section{Introduction}

%It was first shown by Bell \cite{Bell} that statistical results predicted by quantum mechanics (QM) for measurements of spacelike separated parties cannot be reproduced by local realistic (LR) theories. More precisely, Bell showed that joint probability distributions arise from QM which violate inequalities, now called Bell inequalities, which must hold for any LR model. 
%A version called the CHSH inequality \cite{CHSH} allows one in principle to 
%experimentally test QM versus LR. And indeed, experiments \cite{exp1} have shown
%the violation of such Bell inequalities, ruling out (modulo certain loopholes)
%the possibility of finding a LR model alternative to QM, and forcing us to abandon
%Einstein, Podolsky and Rosen's  \cite{EPR} notion of ``elements of reality''.

The violation of Bell inequalities \cite{Bell} by certain quantum correlations can be seen as a nonclassical property of those correlations. This  ``quantum nonclassicality''  has its roots in quantum entanglement. There are several ways to quantify entanglement of which one is the so-called entanglement entropy of a quantum state \cite{Ben96}. 
% [[Choose this option:]]
%Consider a pure state $\ket{\psi}\in\Hc_A\otimes\Hc_B$ with $\Hc_A=\Hc_B=\C^D$, written in its Schmidt decomposition $\ket{\psi}=\sum_{i=0}^{D-1}\la_i\ket{ii}$, where $\ket{ii}\equiv\ket{i}_A\otimes\ket{i}_B$ and $\{i\}$ defines orthonormal bases for $\Hc_A$ and $\Hc_B$. The Schmidt coefficients satisfy  $\la_i\ge0$ and $\sum_{i=0}^{D-1}\la_i^2=1$. Entanglement entropy of a pure state $\ket{\psi}$ is defined as $E(\psi)=-\mathrm{Tr}(\rho_A\log\rho_A)$, where $\rho_A=\mathrm{Tr}_B(\ket{\psi}\bra{\psi})$. In terms of the Schmidt coefficients this reads $E(\psi)=-\sum_{i=0}^{D-1}\la_i^2\log\la_i^2$. The quantum state with the maximum entanglement entropy, the so-called maximally entangled state, is the one with equal Schmidt coefficients $\ket{\Phi}= \sum_{i=0}^{d-1}\ket{ii}/\sqrt{D}$. Such maximally entangled states play an important role in quantum information science \cite{Nielsen}. 
% [[Or this option:]]
Quantum states with maximal entanglement entropy, so-called maximally entangled states, play an important role in quantum information science \cite{Nielsen}. 
It was long believed that the maximally entangled state must also be the ``most nonclassical'' state in the sense of maximal violation of Bell inequalities. Although this is true for the CHSH inequality \cite{CHSH}, it was given evidence in \cite{Ac02,optimal} that this is not true for the more complex CGLMP inequality \cite{CGLMP}, as also exposed in \cite{proceedings}.

In the following, we investigate maximal nonclassicality in the context of the CGLMP.
%, and in the case that the number of possible outcomes is increased indefinitely. 
We present a new simplified version of the CGLMP inequality. As in \cite{Ac02,optimal} numerical analysis suggests that the optimal state for each number of outcomes above $d\!=\!2$ is not maximally entangled, where we mainly work with the assumption that the dimension of the Hilbert space $D$ is equal to the number of outcomes $d$ as in \cite{Ac02,optimal}, but also investigate the case of $d<D$ and the validity of this assumption. We give numerical evidence that the best measurements are the well-known (conjectured) best measurements with the maximally entangled state. The simple form of our new version of CGLMP enables us to effectively extend the numerical search to a number of measurement outcomes and dimension of the Hilbert spaces of the order of $10^6$. We observe that for these large values of $d$ the new version of CGLMP seems to reach its absolute bound at the boundary of the polytope of all probability vectors. This gives numerical evidence for the tightness of a quantum Bell inequality (or generalized Csirelson inequality) for the $2\times2\times\infty$ scenario.

%\section{
{\sl
The $2\times 2 \times d$ Bell scenario and a new version of the CGLMP inequality:}
Let us consider the standard scenario of the CGLMP inequality \cite{CGLMP} which consists of two spacelike separated parties, Alice and Bob. Both share a copy of a pure state $\ket{\psi}\in \C^D\otimes\C^D$ on the composite system. Let Alice and Bob have a choice of performing two different projective measurements which each can have $d$ possible outcomes, where $d\leq D$. We call this a $2\times 2\times d$ scenario. 

Let $A_a^i$, $a=1,2$ and $i=0,..,d-1$ denote the positive operators corresponding to Alice's measurement $a$ with outcome $i$ and similar for Bob, $B_b^j$. They satisfy $\sum_{i=0}^{d-1} A_a^i =\one$. The probability predicted by quantum mechanics (QM) that Alice obtains the outcome $i$ and that Bob obtains the outcome $j$ conditioned on Alice has chosen measurement $a$ and Bob measurement $b$ then reads
\begin{equation}\label{eq:QMprob}
\prob{i,j|a,b}= \mathrm{Tr}\left(A_a^i \otimes B_b^j \ket{\psi}\bra{\psi} \right).
\end{equation}
 
Let us on the other hand consider the framework of local realistic (LR) theories, %. In a local model any correlation between the probability distributions of Alice and Bob must come from initially shared data $\la$ in the region of intersection of Alice's and Bob's past lightcones, their mutual past. Hence, in a LR theory the joint probability distribution can be written as
where the joint probability distribution can be written as
\begin{equation}\label{eq:LR}
\probL{i,j|a,b}= \sum_\la p(\la) \pr{i|a,\la} \pr{j|b,\la},
\end{equation}
meaning that conditioned on their mutual past the probability distributions of Alice and Bob are uncorrelated.

As already mentioned, QM is nonclassical in the sense that there exist joint probability distributions $\prob{i,j|a,b}$ arising from QM which do not admit a local realistic representation in the form of \eqref{eq:LR}. Bell \cite{Bell} was the first to put this statement into a testable form in terms of an inequality which is violated for nonclassical probability distributions.  

We now  give a new Bell inequality for the $2\times 2 \times d$ Bell scenario:

\begin{eqnarray}\label{eq:inequality}
\!\!\!\probL{A_2<B_2}+\probL{B_2<A_1}+\probL{A_1<B_1}+\nonumber\\
+\,\probL{B_1\leq A_2}\geqslant 1,\!\!\!\!\!
\end{eqnarray}
where
%\begin{equation}
$\probL{A_a<B_b}=\sum_{i<j} \probL{i,j|a,b}$.
%\end{equation}

This inequality can be easily proven. Let us start with the following obvious statement
$\{A_2\geq B_2\} \cap \{B_2 \geq A_1\} \cap \{A_1\geq B_1\} \subseteq \{A_2\geq B_1\}$.
Taking the complement we get $\{A_2<B_1\}\subseteq \{A_2<B_2\} \cup  \{B_2<A_1\}\cup \{A_1<B_1\}$. This implies for the probabilities that
%\begin{eqnarray}
$\probL{A_2<B_1} =   1- \probL{A_2 \geq B_1}%\,\,\,\,\,\, \,\,\,\,\,\, \,\,\,\,\,\, \,\,\,\,\,\, \,\,\,\,\,\,\,\,\,\,\,\, \nonumber\\
 \leq   \probL{A_2<B_2} +  \probL{B_2<A_1} +\probL{A_1<B_1}
 $
%\end{eqnarray}
which completes the proof.

The new version $\eqref{eq:inequality}$ of the CGLMP inequality has apart from its simple form several advantages over previous versions. One advantage is that the inequality does not depend on the actual values of the measurement outcomes, only their relative order on the real line matters. For the case of measurements with outcomes $0,...,d-1$ this inequality implies another simplified version of the CGLMP inequality presented in \cite{optimal}, as well as the original CGLMP inequality. Another advantage is that inequality $\eqref{eq:inequality}$  reads the same for all values of $d$. Further, the way the new inequality is derived might be interesting for finding new, simpler inequalities for other Bell settings, such as the $2\times 3\times 2$ Bell setting.

In the following section we will investigate the maximal violation of inequality $\eqref{eq:inequality}$ by QM for large values of the number of outcomes and dimension of the Hilbert space.

\begin{table}[t]
\caption{Violation of the CGLMP inequality}
\begin{center}
\footnotesize
\begin{tabular}{|c|c||c|c|c|c|c|c|c|c|c|c|}
\hline
$d$ & $ \min\Ac $
&$\lambda_0$&$\lambda_1$&$\lambda_2$&$\lambda_3$&$\lambda_4$
\\\hline\hline
2 & $0.7929$ &$0.7071$&$0.7071$&-&-&- \\\hline
3 & $0.6950$ &$0.6169$&$0.4888$&$0.6169$&-&- \\\hline
4 & $0.6352$ &$0.5686$&$0.4204$&$0.4204$&$0.5686$&-\\\hline
5 & $0.5937$ &$0.5368$&$0.3859$&$0.3859$&$0.3859$&$0.5368$ \\\hline
\end{tabular}
\end{center}
\label{table1}
\end{table}

%\section{
{\sl
Violation of the CGLMP inequality for the maximally entangled state:}\label{sec:max}
%
%Let us take the Hilbert space of Alice and Bob to be $d$-dimensional. A state $\ket{\psi}\in\Hc_A\otimes\Hc_B$ can be written in its Schmidt decomposition, $\ket{\psi}=\sum_{i=0}^{d-1}\la_i\ket{ii}$, where $\ket{ii}\equiv\ket{i}_A\otimes\ket{i}_B$ and $\{i\}$ define orthonormal bases in $\Hc_A$ and $\Hc_B$. Further the Schmidt coefficients fulfill the following requirements $\la_i>0$ and $\sum_{i=0}^{d-1}\la_i^2=1$.
%
% sznew 
In the following we will assume that the dimension of the Hilbert space $D$ is equal to the number of outcomes $d$ which we abbreviate as dimension $d$. We will comment on this assumption at the end of this letter, where we also present numerical evidence for the validity of this assumption.
% sznew
For the maximally entangled state, $\ket{\Phi}=\sum_{i=0}^{d-1}\ket{ii}/\sqrt{d}$, it has long been conjectured that the measurements which maximally violate 
%inequality $\eqref{eq:inequality}$ 
the CGLMP inequality
are described by operators $A_a$ and $B_b$ with the following  eigenvectors \cite{Zukowski, CGLMP},
\begin{eqnarray}
\ket{i}_{A,a} & = &\frac{1}{\sqrt{d}} \sum_{k=0}^{d-1} \exp\left(\mathbf{i}\frac{2\pi}{d}k (i+\alpha_a) \right) \ket{k}_A \label{eq:bestmeasA},\\
\ket{j}_{B,b} & = & \frac{1}{\sqrt{d}} \sum_{l=0}^{d-1} \exp\left(\mathbf{i}\frac{2\pi}{d}l (-j+\beta_b) \right) \ket{l}_B \label{eq:bestmeasB},
\end{eqnarray}
where the phases read $\alpha_1=0$, $\alpha_2=1/2$, $\beta_1=1/4$ and $\beta_2=-1/4$, here $\mathbf{i}=\sqrt{-1}$ is the imaginary number.

We evaluate the left-hand-side of inequality $\eqref{eq:inequality}$ for the joint probabilities arising from QM in the case of the maximally entangled state and the just described measurements. For later purposes we will leave the Schmidt coefficients unspecified throughout this calculation and only equate them to $1/\sqrt{d}$ at the end. We use \eqref{eq:QMprob},
%as follows
%\begin{equation}\label{eq:probeq}
%\prob{A_a<B_b}=\sum_{i<j} \mathrm{Tr}\left(A_a^i \otimes B_b^j \ket{\Phi}\bra{\Phi} \right),
%\end{equation}
where the $A_a^i= \ket{i}_{A,a}\bra{i}_{A,a}$ are the projectors on the corresponding eigenspaces defined in \eqref{eq:bestmeasA}--\eqref{eq:bestmeasB} and similarly for $B_b^j$.
We obtain 
\begin{eqnarray}
\Ac_d(\psi)\!\equiv\!\prob{A_2\!<\!B_2}\!+\!\prob{B_2\!<\!A_1}\!+\!\prob{A_1\!<\!B_1}+\nonumber\\
+\,\prob{B_1\!\leq\! A_2}=\sum_{i=0}^{d-1}\sum_{j=0}^{d-1} M_{ij} \la_i\la_j, \,\,\,\,\,\,\,\,\,\,\,\,
\label{eq:invM}
\end{eqnarray}
where the $d\times d$-matrix $M$ can be simplified to
\begin{equation}\label{eq:M}
M_{ij}=2\,\delta_{ij}-\frac{1}{d}\cos^{-1}\left( \frac{(i-j)\pi}{2 d} \right).
\end{equation}

Putting $\la_i=1/\sqrt{d}$, i.e., looking at the maximally entangled state, we obtain for $d=2$, $\Ac_2(\Phi)=(3-\sqrt{2})/2\approx 0.79289$ which corresponds to the maximal violation of the CHSH inequality know from Csirelson's inequality \footnote{The upper bound from Csirelson's inequality \cite{Cir80} is known to be $A=2\sqrt{2}$. Writing out the probabilities one sees that this relates to a lower bound of $(6-A)/4\approx 0.79289$ for the left hand side of our inequality.}.

It is also interesting to look at the conjectured (it is not known that these are the best measurements) maximal violation of $\eqref{eq:inequality}$ with the infinite dimensional maximally entangled state. We get
%\begin{eqnarray}
%\lim_{d\to\infty} \Ac_d (\Phi) &=&  2-\frac{1}{L^2} \int_0^L\int_0^L \!\!\cos^{-1}\left( \frac{\pi}{2L}(x-y) \right) dx dy\nonumber\\
%&=&2-\frac{16 Cat^2}{\pi^2}\approx 0.515\label{eq:asymp}
%\end{eqnarray}
$\lim_{d\to\infty} \Ac_d (\Phi) = 2-16\, \text{Cat}^2/\pi^2\approx 0.515$ where $\text{Cat}$ is
Catalan's constant, 
reproducing the result obtained in \cite{CGLMP} for the original version of the CGLMP inequality.

In this section we described what are believed to be the best measurements for the CGLMP inequality with the maximally entangled state. Though it is often thought that the maximally entangled state $\ket{\Phi}$ represents the most nonclassical quantum state, evidence has been given in \cite{Ac02} and \cite{optimal} that the states which maximally violate inequality \eqref{eq:inequality} are not maximally entangled.
In the following section we provide further evidence for this and investigate several properties of the optimal state especially in the case of very large values of $d$.

%\section{
{\sl
On the maximal violation of the CGLMP inequality:}
In the previous section we described the measurements which in the case of the maximally entangled state appear to give the maximal violation of inequality \eqref{eq:inequality}. However, as mentioned above, it has already been given evidence that in the case of $d\geq3$ the state which causes the maximum violation of the inequality is actually not the maximally entangled state \cite{Ac02,optimal}. 

%Natural questions which arise at this point are: How can the optimal state be described for larger $d$ and are the corresponding best measurements the same as in the case of the maximally entangled state? Further, what is the maximal violation of inequality \eqref{eq:inequality} as $d$ tends to infinity?

%To address the above questions 
In the following we want to optimize the left-hand-side of inequality \eqref{eq:inequality} over all possible measurements and states. 
% sznew I put this back in
For this purpose we assume that the state of Alice's and Bob's composite system is a pure state $\ket{\psi}\in\C^d\otimes\C^d$ and that the measurements $A_a$ and $B_b$ describing Alice's and Bob's measurement are projective and nondegenerate as also considered above. 
%sznew

%In this case the problem of finding the minimal value of the left-hand-side of inequality \eqref{eq:inequality} becomes a nonlinear $(4d^2+d)$-dimensional optimization problem with respect to certain constraints.

For small values of $d$ we can numerically perform the optimization. The results for the first values are summarized in Table \ref{table1}. Shown are the minimal values of the left-hand-side of inequality \eqref{eq:inequality}, denoted by $\min\Ac_d(\psi,A_a,B_b)$, and the Schmidt coefficients of the optimal state for which $\Ac_d(\psi,A_a,B_b)$ reaches its minimum.

One observes that for $d\geq 3$ the optimal state is not maximally entangled. More precisely, as we will see later the entanglement entropy decreases as $d$ becomes bigger. The optimal states 
arising from the numerical optimization over $\Ac_d(\psi,A_a,B_b)$ agree with results obtained in \cite{Ac02}, but differ from the results in \cite{optimal}. That is because in \cite{optimal} the quantity to be optimized was not the CGLMP inequality, but the Kullback-Leibler divergence (relative entropy) which contrary to common belief is not equivalent to the concept of maximal violation of Bell inequalities \cite{statistical}.
%quantifies the average amount of support in favour of QM against LR per trial \footnote{See \cite{statistical} for an application of the Kullback-Leibler divergence to specify the statistical strength of nonlocality proofs.}. It is important to notice, that contrary to common belief, the concept of maximal violation of Bell inequalities and statistical strength of this nonlocality proofs, as measured by the Kullback-Leibler divergence, are not equivalent. This can be further seen by the comparison of the entanglement entropy of the optimal states in both cases, as we will see later.

Closer analysis of the optimal measurements $A_a^i$ and $B_b^j$ shows that even though the optimal state is not the maximally entangled state the best measurements seem to be the best measurements \eqref{eq:bestmeasA} and \eqref{eq:bestmeasB} of the previous case. Further numerical optimizations for higher values of $d$ give strong evidence that this true in general.

If we assume that \eqref{eq:bestmeasA} and \eqref{eq:bestmeasB} are the best measurements for all values of $d$ we can further simplify the optimization. We have already derived in %Sec.~\ref{sec:max} 
Eq.\ \eqref{eq:invM}
that in the case of the measurements  \eqref{eq:bestmeasA} and \eqref{eq:bestmeasB} we can write 
%\begin{eqnarray}
%\Ac_d(\psi)\!\equiv\!\prob{A_2\!<\!B_2}\!+\!\prob{B_2\!<\!A_1}\!+\!\prob{A_1\!<\!B_1}+\nonumber\\
%+\,\prob{B_1\!\leq\! A_2}=\sum_{i=0}^{d-1}\sum_{j=0}^{d-1} M_{ij} \la_i\la_j, \,\,\,\,\,\,\,\,\,\,\,\,
%\end{eqnarray}
$\Ac_d(\psi)=\sum_{i=0}^{d-1}\sum_{j=0}^{d-1} M_{ij} \la_i\la_j$,
where $\ket{\psi}=\sum_{i=0}^{d-1}\la_i\ket{ii}$ and the $d\times d$-matrix $M$ was given in \eqref{eq:M}.

Hence under this assumption, finding the maximal violation of $\eqref{eq:inequality}$ reduces to finding the smallest eigenvalue of the matrix $M$. The corresponding eigenvector $\{\la_i\}_{i=0}^{d-1}$ gives us  the optimal state.

For $d=2,3$ we obtain
%\begin{eqnarray}
$\min\Ac_2  = (3-\sqrt{2})/2$,  %\quad \,\,\,\,\,\,  
with $\vec{\la}  =(1,1)^T/\sqrt{2}$, % \\
and $\min\Ac_3  = (12-\sqrt{33})/9$, % \quad   
with $\vec{\la}  = (1,\gamma,1)^T/(\sqrt{2+\gamma^2})$,
%\end{eqnarray}
and $\gamma=(\sqrt{11}-\sqrt{3})/2$, agreeing with results presented in \cite{Ac02} where violations of the original CGLMP inequality were investigated.

More interesting becomes the search for eigenvectors with minimal eigenvalue for a large number of possible measurement outcomes. Numerical search for those eigensystems is feasible for very large values of $d$ by use of Arnoldi iteration.

\begin{figure}
\begin{center}
\includegraphics[width=3.5in]{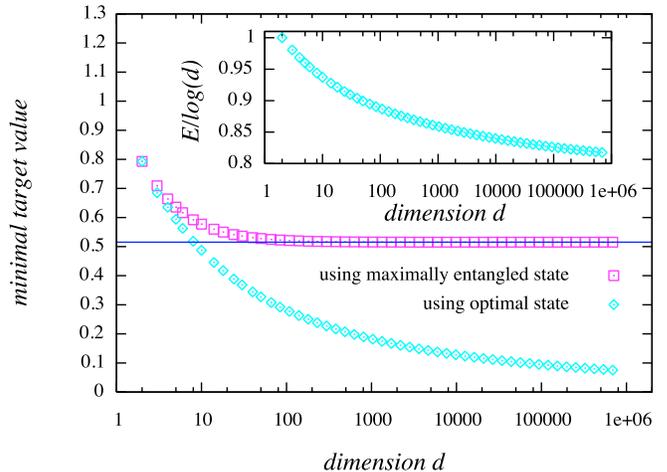}
\caption{Minimal value of the left-hand-side of inequality $\eqref{eq:inequality}$ as a function of the dimension $d$: (i) for the maximally entangled state and (ii) for the optimal state. Inside: Entanglement entropy $E/\log d$ of the optimal state as a function of the dimension $d$.}
\label{fig:min}
\end{center}
\end{figure}

The results of the numerical optimizations are summarized in Fig.~\ref{fig:min}. 
%and Fig.~\ref{fig:entropy}. Fig.~\ref{fig:min} shows 
Shown is the minimal target value, $\Ac_d(\psi)$, as a function of the dimension $d$ for a range from $2$ to $10^6$ both for the case of the maximally entangled state and the optimal state. In the case of the maximally entangled state, $\Ac_d(\Phi)$ approaches very quickly the asymptotic value $\Ac_\infty (\Phi) \approx 0.515$ derived above.%in \eqref{eq:asymp}.

%\begin{figure}[t]
%\begin{center}
%\includegraphics[width=3in]{entropynewsmall}
%\caption{Entanglement entropy $E/\log d$ of the optimal state as a function of the dimension $d$.}
%\label{fig:entropy}
%\end{center}
%\end{figure}

In the case of the optimal state it is interesting that the maximal violation of $\eqref{eq:inequality}$ does not approach an asymptote very quickly. In fact, for very large $d$ it falls off slower than logarithmically with the dimension. The numerical data shown in Fig.~\ref{fig:min} do suggest that the minimal value of $\Ac_d(\psi)$ approaches zero as $d$ tends to infinity. This is very interesting since zero is the absolute minimum of $\Ac_d(\psi)$ on the boundary of the polytope of all probability vectors. If one could show analytically that there exists a optimal state which actually causes $\Ac_d(\psi)$ to approach zero as $d$ tends to infinity, one would have proven a new tight quantum Bell inequality for the $2\times2\times\infty$ scenario (see conjecture 
 at the end of this section).

\begin{figure}[t]
\begin{center}
\includegraphics[width=3in]{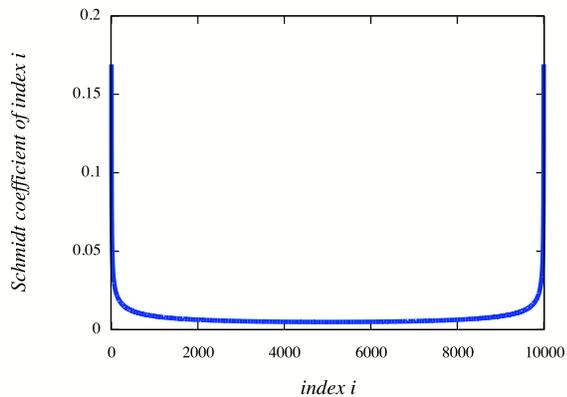}
\caption{The typical shape of a optimal state for $d\geq 3$. Shown are the Schmidt coefficients $\la_i$ of the optimal state for $d=10000$ as a function of the index $i$.}
\label{fig:Ushape}
\end{center}
\end{figure}

Let us now investigate further properties of the optimal states causing the maximal violation of inequality $\eqref{eq:inequality}$. Fig.\ \ref{fig:Ushape} shows the typical shape of a optimal state for $d\geq 3$, namely in the case of $d=10000$. Plotted are the Schmidt coefficients $\la_i$ as a function of the index $i$. The reflection symmetry around $(d-1)/2$ can be easily derived from the specific form of the symmetric kernel $M_{ij}$. As $d$ increases the Schmidt coefficient get more and more peaked at $i=0$ and $i=d-1$.

It is also interesting to look at the entanglement entropy of the optimal state. 
%as shown in Fig.~\ref{fig:entropy}. 
Whereas for the maximally entangled state $E(\Phi)/\log d=1$ for all values of $d$, in the case of the optimal state the entanglement entropy decreases with the dimension.  As in the case of the minimal value of $\Ac_d(\psi)$ the entanglement entropy decreases slower than logarithmically, but we are not able to give an asymptotic bound for it. This is contrary to work presented in \cite{optimal}, where the entanglement entropy seemed to approach the asymptotic value $\lim_{d\to\infty}E(\psi)=\ln d\approx 0.69 \log d$. Again, the disagreement is due to the fact that in the latter the quantity to be optimized was not the CGLMP inequality, but rather the Kullback-Leibler divergence.

From the insights gained in this section we state the following conjecture:

\begin{conj}[Quantum Bell inequality]
For $d\!\to\!\infty$ the minimal value of $\prob{A_2<B_2}+\prob{B_2<A_1}+\prob{A_1<B_1}+\prob{B_1\leq A_2}$ converges to zero, where the best measurements for each $d$ are the ones presented above, \eqref{eq:bestmeasA} and \eqref{eq:bestmeasB}, and the optimal states are of the form 
shown in Fig. \ref{fig:Ushape}.
Hence, 
\begin{eqnarray}
\!\!\!\prob{A_2<B_2}+\prob{B_2<A_1}+\prob{A_1<B_1}+\nonumber\\
+\,\prob{B_1\leq A_2}\geqslant 0\!\!\!
\end{eqnarray}
is a \emph{tight} quantum Bell inequality for the $2\times2\times\infty$ Bell setting.
\end{conj}

The fact that the inequality seems to reach its minimal value given by the probability constraints as $d\to\infty$ also relates to recent results derived in \cite{stefano} for a chained version of the CGLMP inequality.

%A proof of this conjecture is work in progress \cite{progress} and will be hopefully reported soon.

%\section{
{\sl
Conclusion:}
A new version of the CGLMP inequality for the $2\times2\times d$ Bell scenario has been presented. 
%The new inequality, besides being compact and easy to prove, has several attractive features including independence of the number of outcomes and their values as well as its simple form, leading to efficient numerical investigation.
%We briefly reviewed the (conjectured) best measurements in the case of the maximally entangled state for violation of the CGLMP inequality. 
Numerically, under the assumption that the number of outcomes is equal to the dimension of the Hilbert space $D$, the optimal states are not maximally entangled for $d\geq 3$, though the best measurements with respect to those states are the same as for the maximally entangled state.

We investigated the maximal violation of this new inequality for very large numbers of measurement outcomes and dimension of the Hilbert space. We analysed the specific form of the best states and their entanglement entropy. It turned out that for increasing dimension the entanglement entropy of the optimal state decreases, agreeing with the observations made in \cite{optimal,Ac02}. Interestingly, the numerics indicate that the maximal violation of the inequality tends, as the number of measurement outcomes and dimension of the Hilbert space tends to infinity, to the absolute bound imposed by the polytope of probability vectors. We conjectured from this a tight quantum Bell inequality for the $2\times2\times\infty$ Bell scenario. An analytical proof of the tightness of this inequality is work in progress which will hopefully appear soon. 

To justify the above assumption that the dimension of the Hilbert space $D$ is equal to the number of possible outcomes $d$ we also numerically analyzed the case of $d\!<\!D$. In particular, we obtained the minimal target value optimized over Schmidt coefficients and all possible combinations of degenerate measurements $A_1,A_2,B_1,B_2$ for $d\!=\!2,3,4$ with $D\!=\!5$ and over randomly selected degenerate measurements for $d\!=\!2,3,4,5$ with $D\!=\!20$. In all cases the smallest obtained target values agreed with the corresponding minimal target values obtained under the assumption that $D\!=\!d$ as summarized in Table \ref{table1} up to an error of $10^{-3}$. This gives strong evidence for the validity of the assumption that $D\!=\!d$ and suggests that the minimal target values obtained under this assumption are also valid for the case of degenerate projective measurements and POVM measurements which can always be realized as projective measurements on a higher-dimensional Hilbert space due to Naimark's theorem. Further, it strengthens the evidence that the optimal state for $d>2$ is not maximally entangled beyond the analysis of \cite{Ac02,optimal}.

\acknowledgments
%Support through
%ENRAGE (European Network on
%Random Geometry), a Marie Curie Research Training Network in the
%European Community's Sixth Framework Programme, network contract
%MRTN-CT-2004-005616 is kindly acknowledged.
%
%S.Z.~acknowledges support of the Marie Curie Training Network ENRAGE, MRTN-CT-2004-005616. We would also like to thank the referees for interesting comments.
S.Z.~was supported by ENRAGE (MRTN-CT-2004-005616). We thank the referees for interesting comments.

%\bibliographystyle{revtex}
%\bibliography{bell}

\begin{thebibliography}{10}
\providecommand*{\bibinfo}[2]{#2}
\providecommand*{\eprint}[1]{#1}
\providecommand*{\url}[1]{#1}

\bibitem{Bell}
\bibinfo{author}{J.~S. Bell}, \bibinfo{journal}{Physics}
  \bibinfo{volume}{\textbf{1}}, \bibinfo{pages}{195} (\bibinfo{date}{1964}).

%\bibitem{exp1}
%\bibinfo{author}{A.~Aspect}, \bibinfo{author}{P.~Grangier}, and
  %\bibinfo{author}{G.~Roger}, \bibinfo{journal}{Phys. Rev. Lett.}
  %\bibinfo{volume}{\textbf{47}}, \bibinfo{pages}{460} (\bibinfo{date}{1981}).
%\bibitem{exp2}
%\bibinfo{author}{W.~Tittel et~al.}, \bibinfo{journal}{ibid}
  %\bibinfo{volume}{\textbf{81}}, \bibinfo{pages}{3563} (\bibinfo{date}{1998}).
%\bibitem{exp3}
%\bibinfo{author}{G.~Weihs et~al.}, \bibinfo{journal}{ibid}
 % \bibinfo{volume}{\textbf{81}}, \bibinfo{pages}{5039} (\bibinfo{date}{1998}).
%\bibitem{exp4}
%\bibinfo{author}{M.~Rowe et~al.}, \bibinfo{journal}{Nature}
  %\bibinfo{volume}{\textbf{409}}, \bibinfo{pages}{791} (\bibinfo{date}{2001}).

%\bibitem{EPR}
%\bibinfo{author}{A.~Einstein}, \bibinfo{author}{B.~Podolsky}, and
  %\bibinfo{author}{N.~Rosen}, \bibinfo{journal}{Phys. Rev.}
  %\bibinfo{volume}{\textbf{47}}, \bibinfo{pages}{777} (\bibinfo{date}{1935}).

\bibitem{Ben96}
\bibinfo{author}{C.~H. Bennett}, \bibinfo{author}{H.~J. Bernstein},
  \bibinfo{author}{S.~Popescu}, and \bibinfo{author}{B.~Schumacher},
  \bibinfo{journal}{Phys. Rev. A} \bibinfo{volume}{\textbf{53}}(4),
  \bibinfo{pages}{2046} (\bibinfo{date}{1996}), \epfmt{arxiv}{quant-ph/9511030}.

\bibitem{Nielsen}
\bibinfo{author}{M.~A. Nielsen} and \bibinfo{author}{I.~L. Chuang},
  \bibinfo{title}{\emph{Quantum Computation and Quantum Information}}
  (\bibinfo{publisher}{Cambridge U. Press},
  \bibinfo{year}{2000}).

\bibitem{CHSH}
\bibinfo{author}{J.~F. Clauser}, \bibinfo{author}{M.~A. Horne},
  \bibinfo{author}{A.~Shimony}, and \bibinfo{author}{R.~A. Holt},
  \bibinfo{journal}{Phys. Rev. Lett.} \bibinfo{volume}{\textbf{23}}(15),
  \bibinfo{pages}{880} (\bibinfo{date}{1969}).

\bibitem{optimal}
\bibinfo{author}{A.~Acin}, \bibinfo{author}{R.~Gill}, and
  \bibinfo{author}{N.~Gisin}, \bibinfo{journal}{Phys. Rev. Lett.}
  \bibinfo{volume}{\textbf{95}}, \bibinfo{pages}{210402}
  (\bibinfo{date}{2005}), \epfmt{arxiv}{quant-ph/0506225}.

\bibitem{Ac02}
\bibinfo{author}{A.~Acin}, \bibinfo{author}{T.~Durt},
  \bibinfo{author}{N.~Gisin}, and \bibinfo{author}{J.I.~Latorre},
  \bibinfo{journal}{Phys. Rev. A} \bibinfo{volume}{\textbf{65}},
  \bibinfo{pages}{052325} (\bibinfo{date}{2002}), \epfmt{arxiv}{quant-ph/0111143}.
  \bibinfo{author}{J.-L.~Chen}, \bibinfo{author}{C.~Wu},
  \bibinfo{author}{L.C.~Kwek}, \bibinfo{author}{C.H.~Oh}, and \bibinfo{author}{M.-L.~Ge},
  \bibinfo{journal}{Phys. Rev. A} \bibinfo{volume}{\textbf{74}},
  \bibinfo{pages}{032106} (\bibinfo{date}{2006}), \epfmt{arxiv}{quant-ph/050722}.


\bibitem{CGLMP}
\bibinfo{author}{D.~Collins}, \bibinfo{author}{N.~Gisin},
  \bibinfo{author}{N.~Linden}, \bibinfo{author}{S.~Massar}, and
  \bibinfo{author}{S.~Popescu}, \bibinfo{journal}{Phys. Rev. Lett.}
  \bibinfo{volume}{\textbf{88}}(4), \bibinfo{pages}{040404}
  (\bibinfo{date}{2002}), \epfmt{arxiv}{quant-ph/0106024}.
  
\bibitem{proceedings}
\bibinfo{author}{R.~D. ~Gill}, \bibinfo{journal}{IMS Lecture Notes Monograph Series} 
  \bibinfo{volume}{\textbf{55}}, \bibinfo{pages}{135}
  (\bibinfo{date}{2007}),
 \epfmt{arxiv}{math.ST/0610115}. 
 
 \bibitem{Zukowski}
\bibinfo{author}{T.~Durt}, \bibinfo{author}{D.~Kaszlikowski},
  \bibinfo{author}{M.~\.Zukowski},  
  \bibinfo{journal}{Phys. Rev. A}
  \bibinfo{volume}{\textbf{64}}, \bibinfo{pages}{024101}
  (\bibinfo{date}{2001}), \epfmt{arxiv}{quant-ph/0101084}.

\bibitem{Cir80}
\bibinfo{author}{B.~S. Cirel'son}, \bibinfo{journal}{Lett. Math. Phys.}
  \bibinfo{volume}{\textbf{4}}(2), \bibinfo{pages}{93} (\bibinfo{date}{1980}).

\bibitem{statistical}
\bibinfo{author}{W.~van Dam}, \bibinfo{author}{P.~Gr\"unwald}, and
  \bibinfo{author}{R.~Gill}, \bibinfo{journal}{IEEE-Trans.\ Inf.\
  Th.} \bibinfo{volume}{\textbf{51}}, \bibinfo{pages}{2812}
  (\bibinfo{date}{2005}), \epfmt{arxiv}{quant-ph/0307125}.
  
  
  \bibitem{stefano}
\bibinfo{author}{J.~Barrett}, \bibinfo{author}{A.~Kent},
  \bibinfo{author}{S.~Pironio}, \bibinfo{journal}{Phys. Rev. Lett.}
  \bibinfo{volume}{\textbf{97}}, \bibinfo{pages}{170409}
  (\bibinfo{date}{2006}), \epfmt{arxiv}{quant-ph/0605182}.
  
%  \bibitem{progress}
% \bibinfo{author}{R.~Gill}, \bibinfo{author}{J.~Kahn} and
 %\bibinfo{author}{S.~Zohren}, \bibinfo{journal}{in preparation}.


\end{thebibliography}

\end{document}